# PENERAPAN METODE SVM-BASED MACHINE LEARNING UNTUK MENGANALISA PENGGUNA DATA TRAFIK INTERNET

**Muhammad Surahman[1], Leon A. Abdillah[2], Ferdiansyah[3]**
Fakultas Ilmu Komputer, Universitas Bina Darma
Email: muhammadsurahman47@gmail.com[1], leon.abdillah@yahoo.com[2*]

## ABSTRACTS

Internet usage is an important requirement that supports the performance and activities on campus. To control internet usage, it is necessary to know the distribution of internet usage. By utilizing a number of machine learning algorithms and WEKA software, the research is carried out by observation and taking data from wifi hotspots on campus. The classification method using SVM-Based utilizes the classification method owned by Support Vector Machine (SVM). This study aims to classify data on internet usage so that from this classification can be known destination network, protocol and bandwidth that are widely accessed at certain times. Internet traffic data is retrieved through Wireshark software. Whereas data processing and data processing of internet traffic are processed by WEKA. The results showed: 1) UBD internet usage in week I 133,196 users, week II 304,042 users, 2) Use of Destination Network 24,150 and Use of Protocol 37,321, 3) Destination networks that are often addressed are 172.21.206.143 (week I) and 172.21.172.234 (week II), protocols that are often used by TCP, and 4) SVM method is a good data mining method for classifying network packet patterns so as to produce network traffic classification according to destination network and protocol.

## ABSTRAK

Pemakaian internet merupakan kebutuhan yang penting yang mendukung kinerja dan aktivitas di kampus. Untuk mengontrol pemakaian internet maka perlu diketahui sebaran pemakaian internetnya. Dengan memanfaatkan sejumlah algoritma machine learning dan software WEKA maka peneliti melakukan observasi dengan mengambil data dari wifi hotspot di kampus. Metode klasifikasi menggunakan SVM-Based ini memanfaatkan metode klasifikasi yang dimiliki oleh Support Vector Machine (SVM). Penelitian ini bertujuan untuk mengklasifikasi data pemakaian internet sehingga dari klasifikasi tersebut dapat diketahui destination network, protocol dan lebar bandwidth yang banyak diakses pada waktu tertentu. Data trafik internet diambil melalui software Wireshark. Sedangkan pengolahan data dan processing data trafik internet diolah dengan WEKA. Hasil penelitian menunjukkan: 1) Pemakaian internet UBD pada minggu I 133.196 pengguna, minggu II 304.042 pengguna, 2) Penggunaan Destination Network 24.150 dan Penggunaan Protocol 37.321, 3) Destination Network yang sering dituju adalah 172.21.206.143 (minggu I) dan 172.21.172.234 (minggu II), Protocol yang sering digunakan TCP, dan 4) Metode SVM adalah metode data mining yang baik untuk mengklasifikasikan pola-pola paket jaringan sehingga menghasilkan klasifikasi trafik jaringan sesuai dengan destination network dan protocol.

**Kata-kata kunci:** Data Mining, Machine Learning, Network Traffic, Supervised Vector Machine, WEKA.

## 1. PENDAHULUAN

Kemanjuan teknologi informasi (TI) tidak hanya memberikan sejumlah fasilitas yang canggih [1], seperti internet, namun juga perlu dikelola dengan baik agar konsumsi datanya bisa terkendali. Identifikasi akurat dari lalu lintas jaringan (*network traffic*) adalah langkah penting untuk meningkatkan banyak layanan jaringan [2]. Pada era digital sekarang ini hampir setiap aktivitas menggunakan internet. Fasilitas internet merupakan salah satu bagian penting dari infrastruktur [3] kampus pada saat ini. Teknologi *wireless* memungkinkan jaringan internet dengan menggunakan *hotspot*. *Hotspot* sendiri adalah jaringan *wireless* dengan radio frekuensi [4] untuk pertukaran data antara perangkat dengan *access point* (AP). Dengan adanya internet yang handal maka proses pengolahan data, pencarian materi pendukung perkuliahan, proses belajar mengajar secara *online* bisa berjalan dengan baik. Bagian penting dari fasilitas internet adalah besarnya *bandwidth* yang disediakan [5].

Metode klasifikasi data terhadap trafik internet bisa digunakan untuk mengetahui data pemakaian internet yang ada sehingga pengaturan *bandwidth* dapat dilakukan agar tercapai suatu koneksi internet yang handal dan stabil. Penerapan metode *Support Vector Machine* (SVM) bekerja berdasarkan prinsip *Structural Risk Minimization* (SRM) dengan tujuan menemukan *hyperplane* terbaik untuk memisahkan dua buah kelas pada ruang *input*. *Supervised learning* merupakan suatu sebuah pedekatan dimana sudah terdapat data yang dilatih, dan terdapat variabel yang ditargetkan sehingga tujuan dari pendekatan ni adalah mengelompokkan data ke data yang sudah ada. *Supervised learning* memiliki metode klasifikasi maupun regresi.

Universitas Bina Darma (UBD) adalah Perguruan Tinggi Swasta (PTS) yang mengasuh dan mengembangkan ilmu dan keahlian profesional. Universitas Bina Darma saat ini memiliki kapasitas *bandwidth* internet 1Gb/s dan akses ke jalur *inherent* hingga 1Gb/s. Akses internet dan *inherent* tersebut dimanfaatkan untuk menunjang sistem pembelajaran dengan dilengkapi sistem akademis, *e-learning*, dan lain sebagainya. Untuk mempercepat akses informasi Universitas Bina Darma saat ini juga sudah menyediakan layanan *hotspot* yaitu sebuah area dimana pada area tersebut tersedia koneksi internet *wireless* yang dapat diakses melalui *notebook*, PDA maupun perangkat lainnya yang mendukung teknologi tersebut. *Hotspot* tersebut disediakan bagi dosen dan mahasiswa untuk mengakses internet. *Hotspot* di Universitas Bina Darma terdapat beberapa titik area jangkauan yaitu di kampus Utama (hampir seluruh lantai). Untuk pengembangan selanjutnya diharapkan di seluruh lingkungan kampus Universitas Bina Darma terjangkau layanan *hotspot*.

Sejumlah penelitian penulis pelajari sebagai pedoman pada penelitian ini, antara lain: 1) *Network Traffic Classification using Support Vector Machine and Artificial Neural Network* [6], 2) *Network Traffic Classification via Neural Networks* [2], 3) Klasifikasi Data Trafik Internet Menggunakan Metode *Bayes Network* (Studi Kasus Jaringan Internet Universitas Semarang) [5]. Dari uraian diatas peneliti mencoba untuk menerapkan metode klasifikasi SVM untuk mengklasifikasikan jaringan internet mahasiswa Universitas Bina Darma. Sehingga dari klasifikasi tersebut dapat diketahui *destination network*, *protocol* yang banyak diakses.

## 2. METODOLOGI PENELITIAN

### 2.1 Metode Penelitian

Metode yang digunakan dalam penelitian ini adalah *Experiment*. Metode eksperimental merupakan metode penelitian yang memungkinkan seorang peneliti untuk membangun hubungan sebab dan akibat melalui manipulasi variabel dan kontrol situasi [7]. Salah satu metode yang dapat dilakukan di laboratorium maupun di alam terbuka. Metode ini mempunyai arti penting karena selain memberi pengalaman praktis yang dapat membentuk persamaan dan kemauan siswa, metode ini juga melibatkan aktivitas secara langsung.

### 2.2 Metode Pengumpulan Data

Untuk melakukan analisis terhadap trafik internet, maka perlu melakukan pengumpulan data. Data-data yang peneliti perlukan dan dianggap relevan dengan masalah yang peneliti teliti dikelompokkan menjadi 2 (dua) yaitu data primer, dan data sekunder.

Untuk data primer dilakukan dengan melakukan *capturing* ke dalam jaringan wifi internet, observasi, dan wawancara. Sedangkan untuk mendapatkan data sekunder, dilakukan cara mempelajari penelitian terdahulu studi pustaka dan dokumentasi.

**2.3 Machine Learning, Supervised Learning, Support Vertor Machine**

*Machine learning* adalah praktik pemrograman komputer untuk belajar dari data [8]. *Machine learning* merupakan bagian dari *artificial intelligence*. Aplikasi metode-metode *machine learning* untuk *database* yang besar disebut *data mining* [9]. Banyak hal yang dipelajari, akan tetapi pada dasarnya ada 4 (empat) hal pokok yang dipelajari dalam *machine learning*, yaitu: 1) Pembelajaran Terarah (*Supervised Learning*), 2) Pembelajaran Tak Terarah (*Unsupervised Learning*), 3) Pembelajaran Semi Terarah (*Semi-supervised Learning*), dan 4) *Reinforcement Learning*.

Kategori "learning problems" dapat dilihat pada Gambar 1 [10]. Pengelompokan ini didasarkan pada apakah rentang f adalah terbatas (dan tidak berurutan) atau kontinu, dan apakah elemen-elemen rentang diamati secara langsung atau tidak.

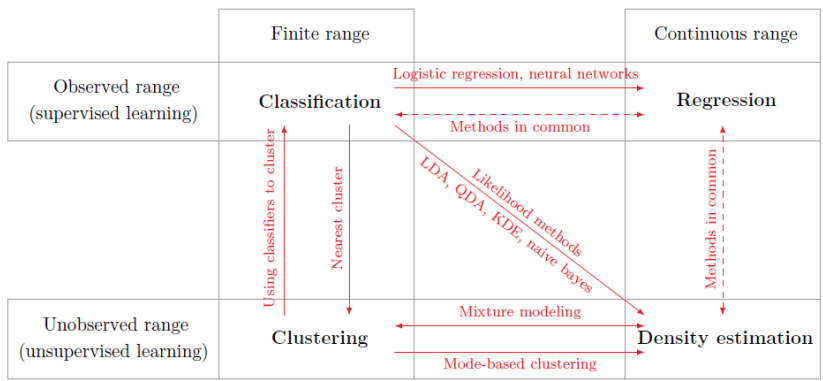

**Gambar 1. Kategori Learning Problems**

Pada penelitian ini menggunakan *supervised learning*. *Supervised learning* merupakan suatu sebuah pedekatan dimana sudah terdapat data yang dilatih, dan terdapat variabel yang ditargetkan sehingga tujuan dari pendekatan ini adalah mengelompokkan data ke data yang sudah ada. *Supervised learning* memiliki metode klasifikasi maupun regresi.

Metode *Support Vector Machine* (SVM) adalah metode *data mining* yang baik untuk mengklasifikasikan pola-pola paket jaringan sehingga menghasilkan klasifikasi trafik jaringan sesuai dengan *destination network* dan *protocol*. SVM pertama kali diperkenalkan oleh Vapnik pada tahun 1992 [11] sebagai rangkaian harmonis konsep-konsep unggulan dalam bidang pengenalan pola.

**2.4 *WEKA***

Proyek WEKA bertujuan untuk menyediakan koleksi lengkap algoritma *machine learning* dan alat pemrosesan data untuk para peneliti dan praktisi [12]. WEKA juga sangat populer sebagai tool untuk data mining dengan machine learning [13].

**2.5 *Flowchart***

Untuk mempermudah pemahaman aktivitas penelitian, maka dibuatkan suatu f*lowchart* yang terdiri atas 6 (enam) aktivitas utama. Aktivitas tersebut, yaitu: 1) *Login* ke *wifi hostspot*, 2) *Capture* dengan Wireshark, 3) *Save dataset* dalam format csv, 4) *Convert dataset* csv tersebut ke format arff, 5) Analisa dengan menggunakan WEKA, dan 6) Gunakan modul klasifikasi dengan *supervised vector machine* (SVM) (Gambar 1).

**Gambar 2. Alur Kerja**

## 3. HASIL DAN PEMBAHASAN

Penelitian ini membahas tentang jaringan internet Universitas Mahasiswa Bina Darma dan melihat aktivitas lalu lintas jaringan komputer. Dengan melihat semua paket-paket data dari setiap protocol maka dapat terlihat segala aktivitas lalu lintas komunikasi data yang selama ini terbungkus dengan rapi. Disini dapat terlihat setiap paket data, baik yang memang merupakan sebuah serangan atau sebuah pemetaan bahkan pengenalan identitas yang akan dituju, sehingga dengan menganalisa setiap paket data tersebut dapat diambil pembelajaran mengenai infrastruktur jaringan.

### 3.1 *Preprocessing*

Pada tahapan Prepocessing ini melakukan *capturing* data atau menangkap lalu lintas jaringan internet mahasiswa Universitas Bina Darma dengan menggunakan *software wireshark*. *Wireshark* merupakan *tool network analyzer* yang mampu menangkap paket-paket data atau informasi yang lalu-lalang pada suatu jaringan. Pada penelitian ini akan masuk terlebih dahulu atau *connect* ke Wi-Fi Mahasiswa UBD.

*Wireshark* akan menampilkan semua informasi tentang *packet* yang keluar dan masuk dalam *interface* yang terpilih. Informasi perbaris pengiriman paket yang direkam, termasuk didalamnya: 1) *No* (pesanan paket diterima), 2) *Time* (waktu paket diterima dari mulai paket dikirim), 3) *Source* (IP Address dari perangkat yang menerima paket), 4) *Destination* (IP address dari perangkat yang menerima paket), 5) *Protocol* (Tipe paket data seperti TCP, UDP, RTP, dsb), 6) *Length* (Panjang Frame dari paket, jika ukuran /size), dan 7) *Info* (Informasi spesifik yang membentuk paket).

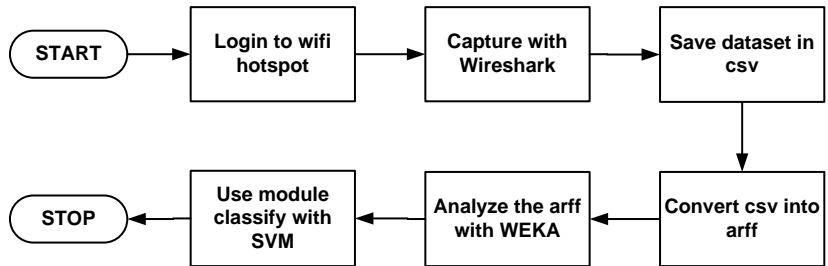

**Gambar 3. Hasil *Capturing Preprocessing***

Setelah melakukan *capture data wireshark* kemudian disimpan dan diekspor dengan format ".csv" (Gambar 4) agar data bisa dibaca dengan *software* WEKA. Hasil dari *capturing* data menggunakan

*Wireshark* (berbentuk .csv) digabungkan perminggu di dalam *excel* sehingga data itu diproses dengan WEKA. Isi dari tangkapan Wireshark akan digabungkan dan dibagi menjadi 4 (empat) minggu.

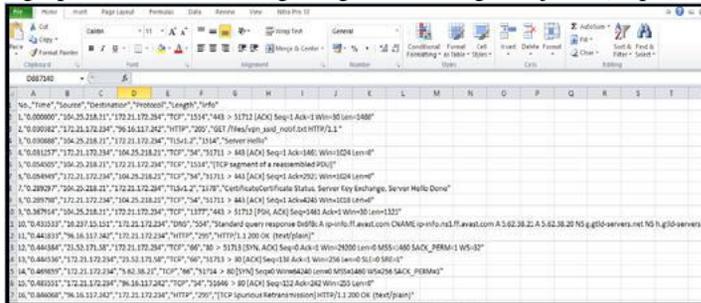

**Gambar 4. Hasil Capture Format csv**

### 3.2 *Postprocessing*

*Post processing* ini adalah proses klasifikasi ini menggunakan *software* WEKA. *Software* Weka [14] merupakan aplikasi yang dibuat dari bahasa pemrograman java yang dapat digunakan untuk membantu pekerjaan *data mining* (penggalian data). WEKA berisi beragam jenis algoritma yang dapat digunakan untuk memproses *dataset* secara langsung atau bisa juga dipanggil melalui kode bahasa java [15].

Data yang diperoleh dari hasil capturing dengan menggunakan wireshark digabungkan menjadi satu file csv perminggu. Data yang berupa file csv diubah format menjadi ".arff" dengan menggunakan WEKA (Gambar 5). WEKA menerima *input* data dalam format *Attribute Relation File Format* (ARFF). Jika menggunakan format file seperti *Comma Separated Values* (CSV) atau *Binary Serialized Instances* (BSI) maka kita harus mengonversi file tersebut mmenjadi format ARFF, begitu juga ketika kita menggunakan format Java kita harus mengubah menjadi format ARFF terlebih dahulu. Format ARFF adalah tipe file teks yang berisi berbagai *instance*. Data yang berhubungan dengan suatu *set* atribut data yang dideskripsikan juga dalam file tersebut.

**Gambar 5. Hasil Capture Format csv**

Data dalam format csv yang di-*capture* dengan Wireshark diubah menjadi format arff. Data langsung berubah menjadi arff langsung di-*save*. Kemudian data yang sudah diubah tersebut dibuka dengan format arff. Data yang sudah diubah menjadi format arff, data dibuka dengan memilih menu *explorer open file* dan buka data dengan format arff.

### 3.3 *Klasifikasi*

Proses klasifikasi ini menggunakan *software* WEKA yang dapat membantu pekerjaan *data mining* (penggalian data). Klasifikasi data terbagi menjadi 7 (tujuh) atribut, yaitu: 1) No, 2) Time, 3) Source, 4) Destination, 5) Protocol, 6) Length, dan 7) Info (Gambar 6). Klasifikasi data dibedakan menjadi 2 (dua) bagian, yaitu: 1) Klasifikasi data destination network, dan 2) Klasifikasi data protocol.

Proses klasifikasi Minggu pertama menggunakan SVM, Proses klasifikasi dengan tombol clasify, kemudian pilih Algoritma klasifikasi dengan klik tombol *choose*. Disini menggunakan libSVM merupakan pendukung berbagai formulasi SVM (Supervised Vector Machine) untuk klasifikasi, regresi, dan estimasi distribsiu.

```
=== Run information ===

Scheme:       weka.classifiers.functions.LibSVM -S 0 -K 2 -D 3 -G 0.0 -R 0.0
Relation:     data 1 Senin-weka.filters.unsupervised.attribute.Remove-R1-3-w
Instances:    133196
Attributes:   3
              Destination
              Protocol
              Length
Test mode:    split 70.0% train, remainder test

=== Classifier model (full training set) ===

LibSVM wrapper, original code by Yasser EL-Manzalawy (= WLSVM)

Time taken to build model: 2305.4 seconds

=== Evaluation on test split ===

Time taken to test model on test split: 124.46 seconds

=== Summary ===

Correctly Classified Instances       37285               93.3081 %
Incorrectly Classified Instances      2674                6.6919 %
Kappa statistic                          0.8687
Mean absolute error                      0.0012
Root mean squared error                  0.0346
Relative absolute error                 12.7104 %
Root relative squared error             50.5043 %
Total Number of Instances            39959
```

**Gambar 7. Contoh Klasifikasi Destination dengan SVM**

Pada ringkasan hasil klasifikasi menunjukkan bahwa jumlah data (*instances*) yang diolah sebanyak 133196 buah. Dari sejumlah data tersebut didapatkan tingkat signifikan datanya sebesar 93,3081% (37285 data) sedangkan tingkat kesalahan .statistik datanya sebesar 6,6919% (2674 data). Dari kedua nilai tersebut dapat disimpulkan bahwa data *trainer* yang digunakan cukup baik.

Pengukuran terhadap kinerja suatu sistem klasifikasi merupakan hal yang penting. Kinerja sistem klasifikasi menggambarkan seberapa baik sistem dalam mengklasifikasikan data. Confusion matrix merupakan salah satu metode yang dapat digunakan untuk mengukur kinerja suatu metode klasifikasi. Pada dasarnya confusion matrix mengandung informasi yang membandingkan hasil klasifikasi yang dilakukan oleh sistem dengan hasil klasifikasi yang seharusnya.

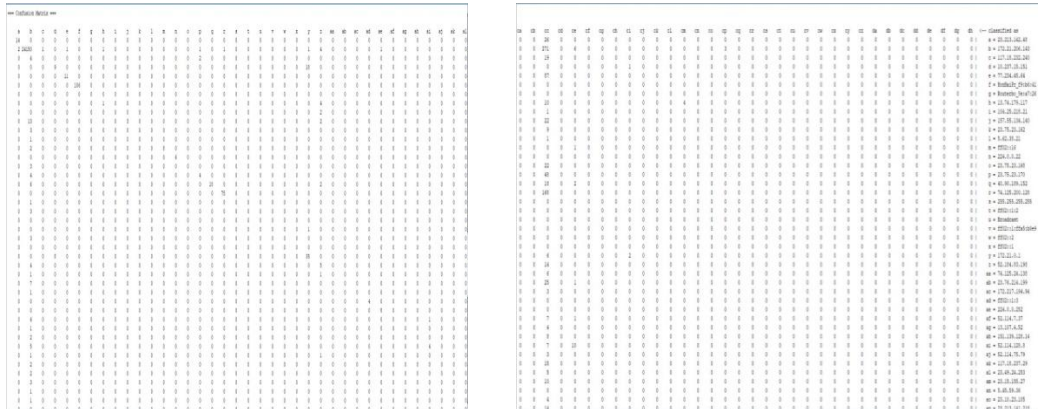

**Gambar 8. Confuction Matrix**

Pada ringkasan hasil klasifikasi menunjukkan bahwa jumlah data (instances) yang diolah sebanyak 133196 buah. Dari sejumlah data tersebut didapatkan tingkat signifikansi datanya sebesar 97,2522% (38861 data) sedangkan tingkat kesalahan statistik datanya sebesar 2,7478% (1098 data). Dari kedua nilai tersebut dapat disimpulkan bahwa data *trainer* yang digunakan cukup baik.

```
Instances:     133196
Attributes:    3
               Destination
               Protocol
               Length
Test mode:     split 70.0% train, remainder test

=== Classifier model (full training set) ===

LibSVM wrapper, original code by Yasser EL-Manzalawy (= WLSVM)

Time taken to build model: 2057.96 seconds

=== Evaluation on test split ===

Time taken to test model on test split: 33.2 seconds

=== Summary ===

Correctly Classified Instances         38861               97.2522 %
Incorrectly Classified Instances        1098                2.7478 %
Kappa statistic                          0.7364
Mean absolute error                      0.0037
Root mean squared error                  0.0605
Relative absolute error                 24.9261 %
Root relative squared error             70.8707 %
Total Number of Instances            39959
```

**Gambar 9. Contoh Klasifikasi Protocol dengan SVM**

*Confuction Matrix Protocol* pada dasarnya mengandung informasi yang membandingkan hasil klasifikasi yang dilakukan oleh sistem dengan hasil klasifikasi yang seharusnya. Terlihat yang Lebih mengakses ke Protocol TCP.

```
=== Confusion Matrix ===

    a     b    c   d    e   f   g   h   i   j   k   l   m   n   o   <-- classified as
  891   456    7   0   43   0   0   0   0   1   0   0   0   0   0 |   a = HTTP
  289 37321    5   0   65   0   0   0   0   0   0   0   0   0   0 |   b = TCP
    0    25  108   0   22   0   0   0   0   0   0   0   0   0   0 |   c = DNS
    0     0    0 214    0   0   0   0   0   0   0   0   0   0   0 |   d = ARP
   51    74    7   0  327   0   0   0   0   0   0   0   0   0   0 |   e = TLSv1.2
    0    12    0   0    3   0   0   0   0   0   0   0   0   0   0 |   f = ICMPv6
    0     6    0   0    0   0   0   0   0   0   0   0   0   0   0 |   g = IGMPv3
    0     0    0   0    1   0   0   0   0   0   0   0   0   0   0 |   h = DHCP
    0     0    0   0    3   0   0   0   0   0   0   0   0   0   0 |   i = DHCPv6
    0     9    0   0    3   0   0   0   0   0   0   0   0   0   0 |   j = LLMNR
    1     0    0   0    2   0   0   0   0   0   0   0   0   0   0 |   k = OCSP
    1     1    0   0    2   0   0   0   0   0   0   0   0   0   0 |   l = TLSv1.1
    0     0    0   0    5   0   0   0   0   0   0   0   0   0   0 |   m = NBNS
    0     1    1   0    0   0   0   0   0   0   0   0   0   0   0 |   n = HTTP/XML
    0     1    0   0    1   0   0   0   0   0   0   0   0   0   0 |   o = NTP
```

**Gambar 10. Contoh Klasifikasi Protocol dengan SVM**

Dari hasil klasifikasi .memperlihatkan jumlah total dari .Protocol yang banyak diakses dalam .jaringan internet terdapat 18 protocol yang digunakan.

**Gambar 11. Atribut Klasifikasi**

Dari hasil klasifikasi juga memperlihatkan nilai statistik dari *Length* atau lebar *bandwidth* yang digunakan.

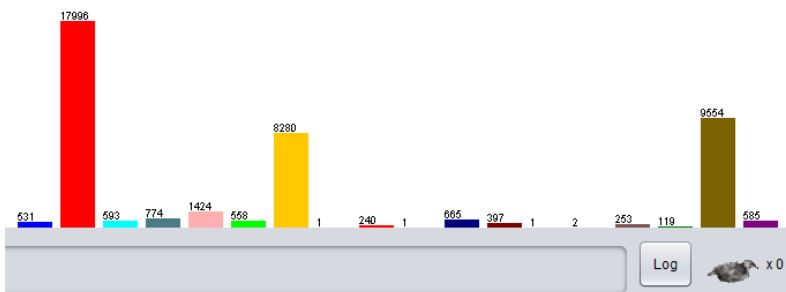

**Gambar 12. Atribut Klasifikasi**

### 3.3 *Hasil Accuracy*

Dari Klasifikasi menggunakan *Supervised Vector Machine* (SVM) Menggunakan software WEKA mendapatkan *Accuracy* yang berbeda di minggu tertantu.

**Tabel 1. Accuracy Data Keseluruhan**

| No | Perminggu | Accuracy | | Persentage Splite |
|---|---|---|---|---|
| | | *Destination* | *Protocol* | |
| 1 | Minggu I | 93.3081% | 97.2522% | 70% |
| 2 | Minggu II | 80.7933% | 96.2144% | 70% |
| 3 | Minggu III | 66.638% | 86.2238% | 70% |
| 4 | Minggu IV | 68.66% | 88.1837% | 70% |

Dari hasil klasifikasi dengan weka penulis mengklsifikasi perminggu dengan nilai accuracy yang berbeda, berdasarkan tabel 4.3 adalah: 1) Klasifikasi Minggu Pertama data Destination dengan Melakukan Splite Testing atau Training set data dengan 70% mendapatkan nilai accuracy 93.3081% dan Ip yang Sering dituju 172.21.206.143 dengan jumlah 24.150 sedangkan Protocolnya Melakukan Splite Testing 70% mendapatkan Nilai Accuracy 97.2522% didapatkan protocol yang sering digunakan TCP 37.321, 2) Minggu Kedua Melakukan Klasifikasi data atau Training set dengan Splite Testing 70% mendapatkan destination 80.7933% dengan IP yang sering dituju 172.21.172.234 sebanyak 60035 dan Protocolnya melakuakan Splite Testing 70% Mendapatkan accuracy data 96.2144% dengan Protocol yang digunakan TCP sebanyak 85.739, 3) Minggu Ketiga Melakukan klasifikasi data atau training set dengan Splite Testing 66% mendapat accuracy destination 66.638%dengan ip yang dituju 172.21.61.73 sebanyak 5.945 dan protocolnya melakukan Splite Testing 70% mendapatkan accuracy data 86.2238% dengan protocol yang digunakan TCP sebanyak 5.282, dan 4) Minggu keempat melakukan Klasifikasi data atau training set dengan Splite Testing 66% mendapat accuracy destination 68.66%dengan ip yang dituju 172.21.61.73 sebanyak 2838 dan Protocol melakukan Splite Testing 70% mendapatkan accuracy data 88.1837% dengan protocol yang digunakan TCP sebanyak 8.744.

Dapat dilihat bahwa nilai accuracy destination yang besar adalah minggu pertama dengan acurracy 93.3081%, dan nilai accuracy Protocol yang besar minggu pertama 97.2522% dengan Splite 70% jadi Splite 70% lebih tinggi accuracynya dibandingkan dengan 66%.

### 3.4 Klasifikasi Keseluruhan

Klasifikasi data keseluruhan dari 4 (empat) minggu dari bulan April dan Mei. Selama 4 (empat) minggu mendapatkan jumlah *IP destination* paling banyak.

**Tabel 2. Accuracy Data Keseluruhan**

| No | Perminggu | Destination | Count |
|---|---|---|---|
| 1 | Minggu I | 172.21.206.143 | 24150 |
| 2 | Minggu II | 172.21.2.156 | 2616 |
| 3 | Minggu III | 209.85.229.236 | 1050 |
| 4 | Minggu IV | 172.21.172.234 | 60035 |
| 5 | Minggu I | 173.194.22.40 | 1960 |
| 6 | Minggu II | 172.21.61.73 | 18344 |
| 7 | Minggu III | 172.21.61.73 | 5945 |
| 8 | Minggu IV | 74.125.130.132 | 249 |
| 9 | Minggu I | 152.118.24.168 | 345 |
| 10 | Minggu II | 172.21.61.73 | 2838 |
| 11 | Minggu III | 23.192.194.247 | 1245 |
| 12 | Minggu IV | 172.21.2.156 | 4675 |

Dari klasifikasi data diatas terdapat tabel-tabel yang banyak diakses perminggu yang dikelaloh oleh WEKA terdapat *destination network* dalam 4 minggu terlihat IP destination yang banyak dituju adalah IP 172.21.61.73 pada minggu ke 2 dengan jumlah 18344.

**Tabel 3. Accuracy Data Keseluruhan**

| No | Perminggu | Protocol | Count |
|---|---|---|---|
| 1 | Minggu I | TCP | 37321 |
| 2 | Minggu I | TLSv1.2 | 327 |
| 3 | Minggu I | HTTP | 891 |
| 4 | Minggu II | HTTP | 1295 |
| 5 | Minggu II | TCP | 85739 |
| 6 | Minggu II | DNS | 329 |
| 7 | Minggu II | ARP | 259 |
| 8 | Minggu III | HTTP | 123 |
| 9 | Minggu III | TCP | 5282 |
| 10 | Minggu III | DNS | 142 |
| 11 | Minggu III | TLSv1.2 | 1252 |
| 12 | Minggu III | QUIC | 2217 |
| 13 | Minggu III | HTTP | 89 |
| 14 | Minggu IV | TCP | 8744 |
| 15 | Minggu IV | TLSv1.2 | 777 |
| 16 | Minggu IV | ARP | 193 |
| 17 | Minggu IV | NBNS | 309 |
| 18 | Minggu IV | LLMNR | 581 |
| 19 | Minggu IV | QUIC | 3487 |
| 20 | Minggu IV | MDNS | 286 |
| 21 | Minggu IV | TCP | 37321 |
| 22 | Minggu IV | TLSv1.2 | 327 |

Dari klasifikasi menggunakan *Supervised Vector Machine* (SVM) menggunakan *software* WEKA mendapatkan *protocol* yang banyak digunakan pada minggu tertentu. Dari tabel diatas bisa dilihat bahwa *protocol* yang sering digunakan adalah TCP, dalam setiap minggu banyak menggunakan protocol TCP terutama minggu kedua dengan jumlah sebanyak 19448 yang menggunakan protocol TCP.

## 4. KESIMPULAN

Berdasarkan uraian yang telah dikemukakan pada bab-bab sebelumnya, maka dapat diambil beberapa kesimpulan dalam mencapai tujuan yang diinginkan. Adapun kesimpulan yang dapat diambil adalah sebagai berikut: 1) Hasil analisa Pemakaian internet Universitas Bina Darma pada minggu pertama sebanyak 133.196 pengguna internet dan minggu kedua 304.042 pengguna, 2) Penggunaan Destination Network di Universitas Bina Darma 24.150 dan Penggunaan Protocol yang sebanyak 37.321, 3) Destination Network yang sering dituju pada minggu pertama 172.21.206.143 dan minggu kedua 172.21.172.234 dan Protocol yang sering digunakan TCP, dan 4) Metode SVM (Supervised Vector Machine) adalah metode data mining yang baik untuk mengklasifikasikan pola-pola paket jaringan sehingga menghasilkan klasifikasi trafik jaringan sesuai dengan destination network dan protocol.

Pada akhir penelitian penulis memberikan beberapa saran untuk penelitian yang akan datang terkait dengan penelitian yang berjudul yang berjudul "Penerapan Metode SVM-Based Machine Learning untuk Menganalisa Pengguna data Trafik Internet (Studi Kasus Jaringan Internet WLAN Mahasiswa Bina Darma)" dan khususnya kepada pihak Universitas Bina Darma: 1) Dapat dikembangkan .metode klasifikasi trafik jaringan .yang lain untuk mendeteksi .gangguan jaringan, dan 2) Metode klasifikasi jaringan dapat dioptimalkan dengan menggabungkan metode lain untuk optimasi data.

## DAFTAR PUSTAKA